\documentclass{egpubl}

\makeatletter
\def\ArxivPreprint{%
\Wileypagenumber%
\def\@oddhead{\small\hfill\upshape\textrm{\p@Wileypagenumber}}%
\def\@evenhead{\small\upshape\textrm{\p@Wileypagenumber}\hfill}%
\def\@oddfoot{}\def\@evenfoot{}%
\def\ps@titlepage{\let\@mkboth\@gobbletwo
 \def\@oddhead{\raisebox{\z@}[8pt][1pt]{\parbox{\textwidth}{\small
   \textit{Preprint}\hfill}}}%
 \def\@oddfoot{}%
 \let\@evenhead=\@oddhead
 \let\@evenfoot=\@oddfoot
 \let\sectionmark=\EmptySectionmark
 \let\subsectionmark=\EmptySubsectionmark
}}
\makeatother
\ArxivPreprint
\usepackage[T1]{fontenc}
\usepackage{dfadobe}

\usepackage{cite}
\BibtexOrBiblatex
\electronicVersion
\PrintedOrElectronic
\hypersetup{pdfsubject={Preprint}, pdfkeywords={4D Gaussian Splatting, dynamic scene reconstruction, Bezier splines}}

\usepackage{graphicx}

\usepackage{egweblnk}
\usepackage{overpic}

\usepackage{amsmath}
\usepackage{amssymb}
\usepackage{amsfonts}
\usepackage{multirow}
\usepackage{float}
\usepackage{colortbl}

\title[GaussianSpline]%
      {GaussianSpline: Adaptive Bézier Splines for Highly Dynamic 4D Gaussian Splatting}

\author[Tom Yachil \& Andrei Sharf]
{\parbox{\textwidth}{\centering
  \begin{tabular}{c@{\hspace{4em}}c}
   Tom Yachil & Andrei Sharf \\[1pt]
   {\small\texttt{yachil@post.bgu.ac.il}} & {\small\texttt{asharf@cs.bgu.ac.il}}
  \end{tabular}}
 \\
 {\parbox{\textwidth}{\centering Ben-Gurion University of the Negev, Israel}}
}

\begin{document}

\teaser{
  \centering
  \includegraphics[width=0.8\textwidth]{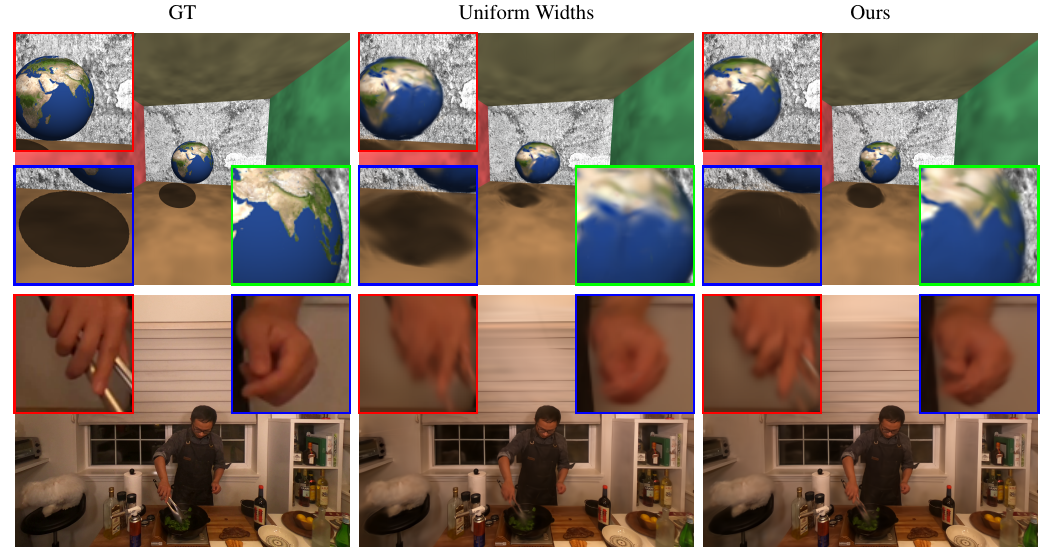}
  \caption{%
Demonstration of GaussianSpline on two highly dynamic 4D scenes (top and bottom rows). Left-to-right (columns) are the ground truth, reconstruction using  splines with uniform-width control points (middle), and our adaptive splines with non-uniform temporal support (right). Zoom-ins highlight that GaussianSpline provides significantly better reconstructions of rapidly changing motion.
  }
  \label{fig:teaser}
}

\maketitle

\begin{abstract}

We present \emph{GaussianSpline}, a dynamic 4D Gaussian Splatting method that models per-Gaussian motion using adaptive, non-uniform piecewise cubic Bézier splines. Unlike approaches based on global deformations, fixed-order polynomials, or uniformly parameterized splines, GaussianSpline jointly learns the Bézier control points and the temporal partition of each Gaussian's trajectory. This adaptive representation allocates greater temporal capacity to highly dynamic regions while maintaining compact trajectories for near-static regions, enabling high-fidelity reconstruction of complex and rapidly changing motion.
To jointly achieve globally smooth motion and local adaptivity, we enforce $G^1$ continuity between consecutive Bézier segments using a reflection-based constraint on their control points. This produces smooth, fully differentiable trajectories while preserving the flexibility of non-uniform temporal parameterization.
Experiments on the Neural 3D Video benchmark, the Google Immersive dataset~\cite{broxton2020immersive}, as well as on synthetic scenes demonstrate that GaussianSpline substantially improves reconstruction quality in scenes with highly non-uniform and challenging dynamics, while maintaining competitive performance on general dynamic scenes compared with state-of-the-art 4D Gaussian Splatting methods.


\end{abstract}

\section{Introduction}
\label{sec:introduction}

Reconstructing dynamic 3D scenes from multi-view video is an emerging problem in computer graphics and vision, with applications in virtual reality, telepresence, and film production. The recent success of 3D Gaussian Splatting (3DGS)~\cite{Kerbl2023_3DGS} has demonstrated that explicit, differentiable representations based on anisotropic Gaussians can achieve high-quality reconstruction and real-time rendering of static 3D scenes, motivating their extension to dynamic 4D content. 
However, moving from static to dynamic Gaussian Splatting introduces a fundamental challenge: the representation must simultaneously capture the spatial structure of the scene and the complex temporal evolution of potentially millions of Gaussians. In particular, accurately modeling highly dynamic regions requires flexible motion trajectories, while avoiding unnecessary temporal degrees of freedom in near-static regions.  

Existing dynamic Gaussian Splatting methods typically rely on neural deformation fields or fixed spatial-temporal representations. Neural deformation approaches augment canonical Gaussians with MLP-based motion predictors~\cite{huang2023sc,Yang2024_Deformable, liang2024gaufregaussiandeformationfields}, introducing additional per-Gaussian computation during rendering. 
Structured 4D approaches instead factorize spacetime using voxel grids or plane-based representations~\cite{Xu2023_4K4D,Wu_2024_CVPR}, but rely on predefined spatial-temporal resolutions. These representations therefore face a trade-off between computational or memory cost and reconstruction fidelity, particularly for scenes with highly non-uniform dynamics, where different regions require different levels of temporal modeling.

To avoid the overhead of neural deformation networks, explicit trajectory-based methods~\cite{li2023spacetime, lin2024gaussian} model Gaussian motion using global parametric functions, such as polynomials or Fourier bases. More recently, piecewise spline representations have been explored for trajectory modeling, offering local control and compact parameterization~\cite{Park_2025_CVPR, yoon2025splinegs}. However, these approaches typically employ a uniform temporal parameterization across Gaussian trajectories, allocating the same temporal capacity regardless of the underlying motion. This limits their ability to efficiently capture the highly heterogeneous dynamics of real-world scenes, where different regions may exhibit slow, smooth motion or rapid and complex changes.

To address these limitations, we introduce \emph{GaussianSpline}, a per-Gaussian trajectory-based 4D Gaussian Splatting framework that models motion using adaptive piecewise cubic Bézier splines with learnable, non-uniform temporal parameterization. Each Gaussian is assigned an independent trajectory initialized with a minimal number of spline segments. During optimization, the Bézier control points and segment intervals are jointly learned, allowing each trajectory to allocate temporal resolution according to its local motion. Rapidly changing motion is captured with finer temporal intervals, while slowly varying motion is represented with broader intervals, avoiding unnecessary temporal degrees of freedom and yielding a compact representation.

To ensure smooth motion across spline segments, we enforce $G^1$ continuity using a reflection-based control-point formulation with direction-locked tangents and learnable magnitude scaling. This yields smooth, fully differentiable trajectories while preserving local flexibility in motion speed and curvature.
GaussianSpline introduces no auxiliary neural networks and integrates directly into the standard 3DGS optimization and densification pipeline, providing a compact and efficient representation for highly non-uniform dynamics.

Experiments on the Neural 3D Video benchmark, the Google Immersive dataset~\cite{broxton2020immersive}, and synthetic scenes demonstrate that GaussianSpline significantly improves reconstruction quality for scenes with complex, highly non-uniform motion, while maintaining competitive performance on general dynamic scenes compared with state-of-the-art 4D Gaussian Splatting methods. Our contributions are summarized as follows:

\begin{itemize}
\item \textbf{Learnable adaptive temporal parameterization.} We jointly learn Bézier control points and non-uniform segment intervals to adapt temporal resolution to each Gaussian's local motion.

\item \textbf{Smooth motion transitions.} We introduce a reflection-based $G^1$ continuity constraint with learnable tangent magnitudes for smooth and flexible trajectory modeling.

\item \textbf{Efficient 4D representation.} Our adaptive trajectories integrate seamlessly into the standard 3DGS framework, providing a compact and efficient representation of dynamic scenes.

\end{itemize}

\section{Related Work}
\label{sec:related_work}

In the following we discuss recent related works on reconstruction of 3D dynamic scenes from multi-view video.

\textbf{Dynamic neural radiance fields.}
Dynamic extensions of NeRF~\cite{mildenhall2020nerf} represent a scene as a time-varying radiance field, either conditioning the field directly on time~\cite{li2022neural3dvideo} or warping each observation into a canonical frame through a learned deformation~\cite{pumarola2021dnerf, park2021hypernerf}, with later work improving robustness on casually captured monocular video~\cite{Liu2023_RoDynRF}. 

These methods produce high-quality free-viewpoint video, but inherit NeRF's volumetric ray marching: rendering a single pixel requires hundreds of network evaluations along a ray, so training takes hours to days and playback remains far from real time~\cite{li2022neural3dvideo}. Grid- and plane-based factorizations~\cite{Fridovich2023_KPlanes} reduce this cost substantially, yet the scene stays an implicit field sampled at a fixed spatial-temporal resolution, so motion is never an explicit quantity that can be allocated where the scene actually moves.

\textbf{Dynamic 3DGS}
The emergence of 3D Gaussian Splatting (3DGS) \cite{Kerbl2023_3DGS} has significantly advanced neural rendering by replacing implicit volumetric sampling with explicit, differentiable rasterization of anisotropic Gaussians. 
Building on this success, recent work has focused on extending 3DGS to dynamic scenes by modeling temporal deformation of Gaussian attributes \cite{luiten2023dynamic, li2023spacetime, Wu_2024_CVPR}.

Luiten et al.~\cite{luiten2023dynamic} directly optimize per-timestep Gaussian positions and rotations under local rigidity constraints, enabling dense 6-DOF tracking without an explicit deformation model. In contrast, MLP-based approaches learn continuous deformation fields in a canonical space. Deformable 3DGS \cite{Yang2024_Deformable} maps Gaussians from canonical to observation space via a neural deformation network, while GauFRe \cite{liang2024gaufregaussiandeformationfields} introduces a forward-warping formulation with an explicit static component to decouple moving and stationary regions. Although expressive, these methods require per-Gaussian MLP evaluations at render time, leading to significant computational overhead \cite{li2023spacetime}. SC-GS \cite{huang2023sc} mitigates this issue by combining deformation networks with sparse spatial constraints to reduce inference cost.

Another line of work improves efficiency through structured spacetime factorization. Originating from NeRF-based representations such as K-Planes \cite{Fridovich2023_KPlanes}, these methods decompose the 4D domain into low-dimensional grids or plane representations. This idea has been adopted in 3DGS extensions such as 4DGS \cite{Wu_2024_CVPR} and 4K4D \cite{Xu2023_4K4D}, which model dynamic Gaussian attributes using structured grid-based embeddings. While computationally efficient, these methods are inherently limited by discrete resolution, leading to a trade-off between memory usage and reconstruction fidelity \cite{Wu_2024_CVPR, li2023spacetime}.

\textbf{Explicit trajectory modeling.}
More recent methods abandon neural deformation networks in favor of explicit motion trajectories. 

Spacetime Gaussians (STG) \cite{li2023spacetime} represent Gaussian motion using polynomial trajectories, improving temporal smoothness and efficiency. Gaussian-Flow \cite{lin2024gaussian} further extends this idea by combining polynomial and Fourier-based representations in a dual-domain formulation, eliminating the need for neural networks while improving expressiveness.

However, fixed-degree polynomial models assign uniform temporal capacity to all Gaussians and may struggle to represent complex, non-linear motion patterns \cite{li2023spacetime}. This limitation has motivated spline-based representations as a more flexible alternative.

\begin{figure*}[t]
  \centering
  \includegraphics[width=\textwidth]{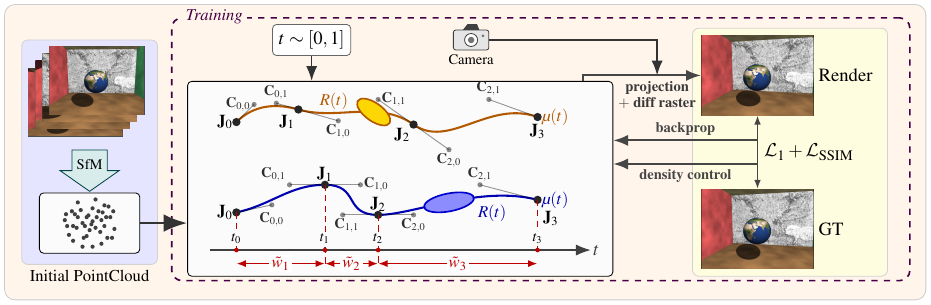}
  \caption{\textbf{Pipeline overview.} Left: multi-view video frames sampled at random timestamps and the initial SfM point cloud. Middle: three per-Gaussian piecewise cubic Bézier trajectories with learnable, non-uniform temporal intervals. Right: Gaussians evolve along their learned trajectories to reconstruct the 4D scene, with rasterized frames compared against ground truth and the resulting photometric loss back-propagated to the spline parameters and Gaussian attributes.}
  \label{fig:pipeline}
\end{figure*}

Spline-based formulations provide smooth and locally controllable representations of motion trajectories. SplineGS \cite{Park_2025_CVPR} introduces piecewise cubic Hermite splines with motion-adaptive control-point pruning, while also reducing reliance on pre-computed camera parameters. However, its formulation targets monocular video sequences rather than multi-view dynamic 4D reconstruction, as considered in our work.

More closely related to our setting, Yoon et al.~\cite{yoon2025splinegs} propose a method, also termed SplineGS, that employs NURBS to represent Gaussian trajectories as combinations of learned representative motion bases, improving temporal smoothness and spatial coherence in multi-view 4D dynamic scenes.

However, both approaches rely on uniform spline parameterizations with fixed control-point distributions. In contrast, our approach jointly learns adaptive, locally refined control-point distributions based on motion complexity. This allows temporal capacity to be allocated where it is most needed, providing a more efficient representation of highly non-uniform motion and improving reconstruction quality with fewer Gaussian splats.

\section{Method}
\label{sec:method}

We first review the standard 3D Gaussian Splatting representation and then extend it to dynamic scenes by assigning each Gaussian a learnable piecewise cubic Bézier trajectory. Unlike fixed temporal parameterizations, our formulation jointly optimizes the spline parameters and the temporal placement of control points, allowing each Gaussian to adapt its temporal resolution to the underlying motion (see Figure~\ref{fig:pipeline}).

\textbf{3D Gaussian Splatting}
3D Gaussian Splatting (3DGS) represents a scene as a set of anisotropic Gaussians, each parameterized by its position $\boldsymbol{\mu}$, covariance matrix $\boldsymbol{\Sigma}$, opacity $o$, and view-dependent color $\mathbf{c}$ represented using spherical harmonics. Each Gaussian is defined as:

\begin{equation}
G(\mathbf{x}) = \exp\left(-\frac{1}{2}(\mathbf{x}-\boldsymbol{\mu})^\top \boldsymbol{\Sigma}^{-1}(\mathbf{x}-\boldsymbol{\mu})\right),
    \label{eq:gaussian}
\end{equation}

with covariance parameterized as:

\begin{equation}
\boldsymbol{\Sigma} = \mathbf{R}\mathbf{S}\mathbf{S}^{\top}\mathbf{R}^{\top},
     \label{eq:covariance}
\end{equation}

where $\mathbf{S}$ and $\mathbf{R}$ denote learnable scaling and rotation matrices, respectively.

For rendering, Gaussians are projected onto the image plane and rasterized using differentiable alpha compositing, where Gaussians are accumulated in depth order:

\begin{equation}
\mathbf{C}=\sum_{i\in N}\mathbf{c}_i\alpha_i\prod_{j=1}^{i-1}(1-\alpha_j),
\end{equation}

Starting from an SfM point cloud, all Gaussian parameters are optimized end-to-end through differentiable rendering, while adaptive density control dynamically splits, clones, and prunes Gaussians during training.
We refer  to~\cite{Kerbl2023_3DGS} for a detailed description of the 3DGS formulation.

\textbf{4D Gaussian Splatting for dynamic scenes.}
To reconstruct a dynamic 4D scene, we lift static Gaussian splats to a time-varying representation over a normalized timeline $t\in[0,1]$. %
In our formulation, only the Gaussian position $\boldsymbol{\mu}(t)$ and rotation $\mathbf{R}(t)$ are time-dependent, while the scale $\mathbf{s}$, opacity $o$, and color $\mathbf{c}$ are shared across all timestamps. The time-varying covariance is therefore given by:
\begin{equation}
\boldsymbol{\Sigma}(t)=\mathbf{R}(t)\mathbf{S}\mathbf{S}^{\top}\mathbf{R}(t)^{\top},
\label{eq:covariance_time}
\end{equation}
and the Gaussian at time $t$ is obtained by evaluating Eq.~\eqref{eq:gaussian} with $\boldsymbol{\mu}(t)$ and $\boldsymbol{\Sigma}(t)$.

This design reflects the observation that many real-world motions can be approximated as piecewise locally rigid deformations, with temporal variation primarily captured by translation and rotation. Restricting temporal variation to these degrees of freedom yields a compact yet flexible representation for dynamic scene modeling.

\textbf{Non-uniform Cubic B\'{e}zier Spline Trajectories}
To model the motion trajectory of a Gaussian splat over time, we use a piecewise cubic Bézier spline defined by $K+1$ joints $\{\mathbf{J}_0, \mathbf{J}_1, \dots, \mathbf{J}_K\}$, $2K$ interior control points $\{\mathbf{C}_{k,0}, \mathbf{C}_{k,1}\}_{k=0}^{K-1}$, and $K$ non-uniform interval widths $\{\mathbf{w}_0, \mathbf{w}_1, \dots, \mathbf{w}_{K-1}\}$. 
Note that all spline parameters are differentiable and jointly optimized within our framework (see Figure~\ref{fig:pipeline} (middle)).

To enforce monotonic partition of the entire temporal domain based on the learned adaptive non-uniform intervals we define:
\begin{equation}
    t_k \;=\; \sum_{j=0}^{k-1} \tilde{w}_j,
    \label{eq:trainable_widths}
\end{equation}

where $\tilde{w}_k \;=\; \frac{\exp(\mathbf{w}_k)}{\sum_{j=0}^{K-1} \exp(\mathbf{w}_j)}$ is segment $k$'s interval width, $t_k$ is $\mathbf{J}_k$'s timestamp, and the joints timestamp sequence $0 = t_0 < t_1 < \dots < t_K = 1$ is a valid monotonic partition of $[0,1]$ by construction.

To evaluate the trajectory at a global timestamp $t \in [0,1]$, we locate the active segment $k$ such that $t_k \le t < t_{k+1}$ and compute the local parameter $u \;=\; \frac{t - t_k}{\tilde{w}_k}$ which maps $u$ to $[0, 1)$ within the active segment. The position $\boldsymbol{\mu}(t)$ is then obtained by evaluating the cubic B\'{e}zier curve on segment $k$'s points, given by:
\begin{equation}
    \mathbf{\mu}(t) = (1-u)^3 \mathbf{J}_k + 3(1-u)^2 u\, \mathbf{C}_{k,0} + 3(1-u) u^2 \mathbf{C}_{k,1} + u^3 \mathbf{J}_{k+1}
    \label{eq:bezier}
\end{equation}

\begin{figure}[t]
  \centering
  \setlength{\tabcolsep}{1.5pt}
  \renewcommand{\arraystretch}{1}
  \begin{tabular}{cc}
  \includegraphics[width=0.5\columnwidth]{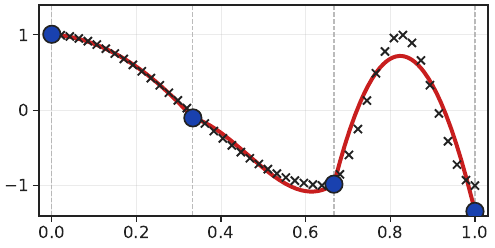} &
  \includegraphics[width=0.5\columnwidth]{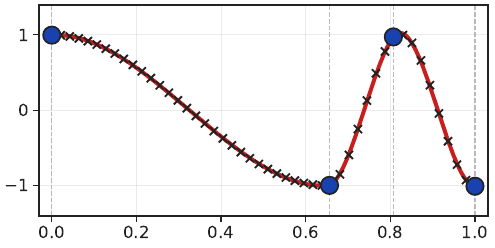} \\
  \end{tabular}
     \caption{Real-data reconstruction of a 1D curve using uniform (left) and our adaptive non-uniform splines (right). Our method jointly learns the temporal interval placement, allocating more control points to high-curvature regions and fewer to slowly varying regions. This adaptive parameterization yields a significantly more accurate reconstruction (red curve) of the target samples (black dots).}
  \label{fig:trainable_widths_toy}
\end{figure}

\textbf{Learnable interval widths.}
Rather than using a uniform partition of the time domain $[0,1]$, we parameterize each spline using learnable interval-width logits $\{\mathbf{w}_0,\mathbf{w}_1,\dots,\mathbf{w}_{K-1}\}$, which define a non-uniform temporal segmentation. Initialized to zero, these logits yield a uniform partition at the beginning of training and are subsequently optimized jointly with the spline joints and control points.

This allows each Gaussian to adapt its temporal resolution to the underlying motion, allocating finer intervals to rapidly changing motion and broader intervals to slowly varying motion, thereby improving reconstruction fidelity over an equivalent uniformly parameterized spline (see Figure~\ref{fig:trainable_widths_toy}).

\begin{figure*}[!t]
  \centering
  \includegraphics[width=\textwidth]{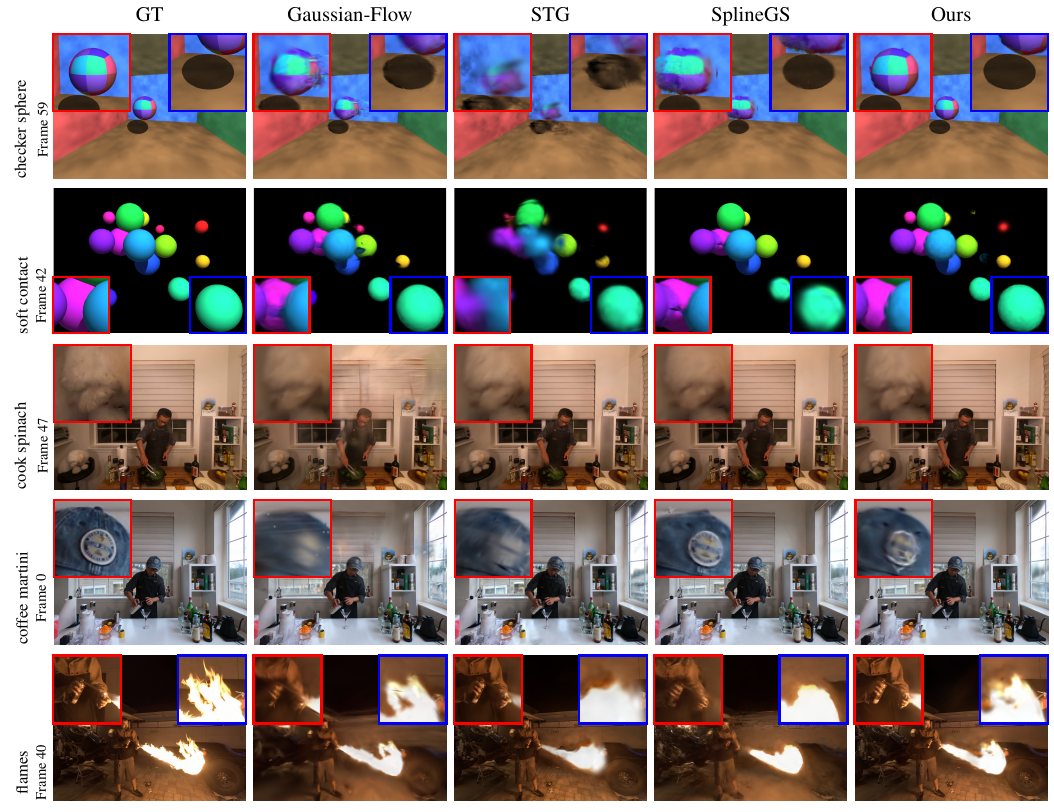}
  \caption{\textbf{Qualitative evaluation} on five dynamic scenes (rows): \emph{checker sphere} (frame 59), \emph{soft contact} (frame 42), \emph{cook spinach} (frame 47), \emph{coffee martini} (frame 0), and \emph{flames} (frame 40), with zoom-ins. From left to right: ground truth, Gaussian-Flow, STG, SplineGS and our method. Zoom-ins demonstrate that our adaptive non-uniform splines better preserve fine details in moving objects while maintaining sharpness in near-static regions, whereas competing methods produce motion blur or lose fine details.
}
  \label{fig:qualitative}
\end{figure*}

\textbf{Learnable motion speed.}
$C^0$ continuity is ensured by shared joints between adjacent segments. Rather than enforcing the stricter $C^1$ continuity, which couples both the direction and speed of motion across neighboring segments through continuity of the first derivative, we impose $G^1$ continuity. This preserves the tangent direction at each internal junction while allowing its magnitude to vary through a learnable positive scalar $\lambda_k > 0$:

\begin{equation}
    \mathbf{C}_{k,0} \;=\; \mathbf{J}_{k} \,+\, \lambda_{k}\,\big(\mathbf{J}_{k} - \mathbf{C}_{k-1,1}\big), \qquad k = 1, \dots, K-1.
    \label{eq:g1_constraint}
\end{equation}

The per-junction scales $\lambda_k$ are initialized to $1$, recovering $C^1$ continuity at the start of training, and are jointly optimized with the spline parameters. This relaxation allows motion speed to vary smoothly across segments while preserving kink-free trajectories and tangent continuity throughout the spline.

\textbf{Per-Gaussian polynomial rotations.}
To compute the per-Gaussian time-dependent rotation matrix $R(t)$, we follow \cite{li2023spacetime} and parameterize the raw quaternion as a first-degree polynomial, mapping its normalized form to the rotation group:
\begin{equation}
q(t) = q_0 + q_1 t, \quad R(t) = \mathcal{R}\!\left(\frac{q(t)}{\|q(t)\|}\right)
\end{equation}
where $q_0$ and $q_1$ denote the base quaternion and its linear velocity, respectively, and $\mathcal{R}(\cdot)$ is the standard mapping from unit quaternions to rotation matrices.

\textbf{Training.}
In each training iteration, we pick a random  frame index from the training set, calculate its timestamp and select a subset of camera views captured at this timestamp. 
Then, every Gaussian is evaluated at this timestamp through its active segment (Eq.~\ref{eq:bezier}), and the photometric loss w.r.t. the selected views is back-propagated to the spline and Gaussian parameters.

Together with the static initialization (all joints and control points set to $\mathbf{x}_0$), this generates the Gaussian motion trajectories. Note that in near-static regions the gradients average or are near zero, leaving the trajectory approximately flat and consuming no temporal capacity.

Furthermore, we apply Gaussians adaptive densification steps, i.e.,  split/clone/prune operations in early stage training iterations (as in~\cite{Kerbl2023_3DGS}).
Afterwards, the set of Gaussians remains fixed, and only the spline parameters (joints, control points, widths, $G^1$ scales) and the Gaussian attributes (opacity, color, covariance) are optimized. 

\section{Experiments and Results}
\label{sec:experiments}

\textbf{Implementation Details.}
We train for 30k iterations using Adam on a single RTX 4090, optimizing an $\mathcal{L}_1$ loss with a D-SSIM term ($\lambda=0.2$), SH degree 3, and batch size 2. Following 3DGS, the position learning rate decays exponentially from $1.6\!\times\!10^{-4}$ to $1.6\!\times\!10^{-6}$, while the learning rates for features, opacity, scaling, and rotation are $2.5\!\times\!10^{-3}$, $5\!\times\!10^{-2}$, $5\!\times\!10^{-3}$, and $10^{-3}$, respectively. The spline joint-width learning rate is set to $0.3\times$ the position learning rate. Densification is performed every 100 iterations from iteration 500 to 9k, using a gradient threshold of $2\!\times\!10^{-4}$ and a maximum of six densification events. Opacity is reset every 3k iterations, and Gaussians with opacity below $0.005$ or scale above $0.1\times$ the scene extent are pruned. Unless otherwise stated, we use $K=3$ spline segments with non-uniform joint spacing and $G^1$ continuity, initialized from approximately 8k COLMAP SfM points from a single frame on N3DV.

\textbf{Baselines}
We compare our method with three other SOTA methods: Gaussian-Flow~\cite{lin2024gaussian}, STG (Spacetime Gaussian) ~\cite{li2023spacetime} and  SplineGS~\cite{yoon2025splinegs}. 

We select these methods because they target the same multi-view dynamic 4D reconstruction setting and, like our approach, explicitly model Gaussian motion trajectories. They therefore provide relevant baselines for evaluating different approaches to temporal motion representation. STG represents Gaussian motion using polynomial trajectories, Gaussian-Flow combines polynomial and Fourier-based representations, and SplineGS employs piecewise cubic Hermite splines with motion-adaptive control-point pruning. 
These methods constitute the current set of SOTA approaches that explicitly model Gaussian motion trajectories in the multi-view dynamic 4D reconstruction setting, providing direct baselines for evaluating our adaptive temporal parameterization on highly non-uniform motion.

In our experiments, we use the default training schedules of 30K, 30K, and 20K iterations for Gaussian-Flow, STG, and SplineGS, respectively. Initialization varies by scene type: for real-world sequences, we initialize from SfM, as ground-truth geometry is unavailable, while for synthetic scenes, we initialize from the ground-truth point clouds. All methods use the same views, frames, and image resolution, and are evaluated at their respective final checkpoints.

For STG, we use the \emph{without temporal opacity} configuration defined in~\cite{li2023spacetime}, in which each Gaussian's opacity remains constant over time. This setting enables a direct comparison of the explicit Gaussian trajectories and motion. Allowing temporal opacity could otherwise represent motion implicitly by fading Gaussians in and out rather than explicitly moving them through space. Since trajectory-based methods model motion through explicit Gaussian trajectories, we retain constant opacity for a consistent comparison.

\textbf{Datasets.}
The Neural 3D Video (N3DV) dataset~\cite{li2022neural3dvideo} comprises multi-view recordings of dynamic real-world scenes captured by 21 synchronized cameras at a resolution of $2704\times2028$ and 30 FPS. The six publicly available scenes contain approximately 10 seconds of video and feature complex non-rigid motion. Following prior work~\cite{Wu_2024_CVPR, lin2024gaussian}, we downsample the images to $1352\times1014$.

We evaluate our method also on the  Google Immersive light field video dataset~\cite{broxton2020immersive} (flame scene), captured by an outward-facing rig of 46 synchronized fisheye cameras mounted on a hemispherical dome. We use the 45 calibrated cameras, holding out \texttt{camera\_0001} for evaluation and training on the remaining 44, over 50 frames. Following STG~\cite{li2023spacetime}, the fisheye images are undistorted to a pinhole model and downsampled to $1280\times960$. The scene features a flamethrower whose rapidly evolving, semi-transparent flame contrasts with a largely static background.

We additionally evaluate on two ray-traced synthetic dynamic scenes, \emph{checker sphere} and \emph{soft contact}, with known ground-truth geometry. Both scenes use the same camera rig of 20 static cameras spanning $90^\circ$ in azimuth across two rings, with 60 frames captured at 30 FPS. Images are rendered at $2704\times2028$ and downsampled to $1352\times1014$, with the central view held out for evaluation.

\emph{checker sphere} features a checker-patterned sphere moving laterally within a five-plane room, following two sinusoidal motion segments with different frequencies. 

\emph{soft contact} contains 16 balls with different masses ($m\propto r^3$) undergoing collisions with an invisible box. Contacts are modeled using a penalty spring-damper over a finite $0.2$\,s interval, producing continuous velocities with large, localized accelerations. Both scenes are procedurally generated, providing per-frame ground-truth geometry for quantitative evaluation.

\begin{table*}[t]
\centering
\caption{\textbf{Quantitative evaluation} on the two synthetic scenes (\emph{checker sphere} and \emph{soft contact}, and the three real-world scenes (\emph{cook spinach} and \emph{coffee martini} from N3DV, and \emph{flames} from Google Immersive~\cite{broxton2020immersive}). We compare our method with three other SOTA: Gaussian-Flow, STG and SplineGS using PSNR, SSIM and LPIPS error metrics.  The best result per scene and metric is shown in \textbf{bold}; \underline{underlined} entries marks the second best. }
\resizebox{\textwidth}{!}{%
\begin{tabular}{lccc|ccc|ccc|ccc|ccc}
\hline
\multirow{2}{*}{Method}
 & \multicolumn{3}{c|}{\emph{checker sphere}}
 & \multicolumn{3}{c|}{\emph{soft contact}}
 & \multicolumn{3}{c|}{\emph{cook spinach}}
 & \multicolumn{3}{c|}{\emph{coffee martini}}
 & \multicolumn{3}{c}{\emph{flames}} \\
\cline{2-4} \cline{5-7} \cline{8-10} \cline{11-13} \cline{14-16}
 & PSNR$\uparrow$ & SSIM$\uparrow$ & LPIPS$\downarrow$
 & PSNR$\uparrow$ & SSIM$\uparrow$ & LPIPS$\downarrow$
 & PSNR$\uparrow$ & SSIM$\uparrow$ & LPIPS$\downarrow$
 & PSNR$\uparrow$ & SSIM$\uparrow$ & LPIPS$\downarrow$
 & PSNR$\uparrow$ & SSIM$\uparrow$ & LPIPS$\downarrow$ \\
\hline
Gaussian-Flow
 & 37.42 & 0.9875 & 0.0495
 & \textbf{31.10} & 0.9829 & 0.0308
 & 24.65 & 0.8771 & 0.2293
 & 23.42 & 0.8571 & 0.2398
 & 21.69 & 0.7453 & 0.3878 \\
STG
 & 34.78 & 0.9920 & 0.0198
 & 23.42 & 0.9623 & 0.0551
 & 31.93 & 0.9516 & \textbf{0.1393}
 & 26.80 & 0.9066 & \textbf{0.1628}
 & 23.96 & 0.8888 & 0.1680 \\
SplineGS
 & 40.37 & 0.9953 & 0.0375
 & 24.55 & 0.9805 & 0.0336
 & 32.67 & \textbf{0.9528} & 0.2007
 & 27.73 & 0.9059 & 0.2188
 & 21.39 & 0.7309 & 0.4742 \\
\textbf{Ours}
 & \textbf{41.18} & \textbf{0.9956} & \textbf{0.0151}
 & \underline{26.84} & \textbf{0.9845} & \textbf{0.0289}
 & \textbf{32.92} & \underline{0.9522} & \underline{0.1531}
 & \textbf{27.99} & \textbf{0.9087} & \underline{0.1743}
 & \textbf{27.42} & \textbf{0.9068} & \textbf{0.1498} \\
\hline
\end{tabular}%
}
\label{tab:all}
\end{table*}

\textbf{Qualitative  Evaluation.}
Figure~\ref{fig:teaser} compares between reconstruction with uniform-width splines (middle) and our adaptive splines with non-uniform temporal support (right). The top and bottom rows show \emph{checker sphere} and \emph{cook spinach}, respectively. The zoomed-in regions demonstrate the benefit of adaptive temporal allocation: GaussianSpline better recovers the earth's sphere-geometry, texture, and shadow (top), as well as the hands, fingers, and gripper (bottom), while uniform splines produce blurred or incomplete details.

Figure~\ref{fig:qualitative} provides qualitative comparisons  on five dynamic scenes (rows): two synthetic sequences with known ground truth 3D motion, \emph{checker sphere} (frame 59) and \emph{soft contact} (frame 42), and three real-world sequences: \emph{cook spinach} (frame 47) and \emph{coffee martini} (frame 0) from N3DV, and \emph{flames} (frame 40) from Google Immersive. Columns show the ground truth, Gaussian-Flow,  STG, SplineGS, and ours with  zoom-ins. 

On the~\emph{checker sphere}, we focus on a rapidly horizontally moving sphere with a checker pattern. Gaussian-Flow blurs the quadrant boundaries, STG completely distorts the sphere and its shadow, while SplineGS blurs the facets and completely misses the red facet. In contrast, GaussianSpline preserves the quadrant boundaries, sphere geometry, and elliptical shadow,  closely matching the ground truth.

\emph{soft contact} represents the most challenging case, with rapid collisions between small spheres. Gaussian-Flow fails to preserve the boundaries of closely colliding spheres, producing bleeding artifacts between them. STG collapses colliding spheres into unified smears, while SplineGS produces inter-sphere artifacts and blur. In contrast, GaussianSpline preserves the individual sphere boundaries and geometry.

The \emph{cook spinach} real-world scene contains regions with different motion patterns. The hand and grippers move continuously, while the dog remains largely static before turning its head in the second half of the sequence. Zoom-ins show that all other methods blur the dog's fur, whereas ours produces the sharpest reconstruction. This improvement comes from our adaptive partition, which concentrates temporal resolution around the localized head motion rather than allocating capacity uniformly across the sequence.

In \emph{coffee martini}, all other methods fail to accurately reconstruct the round cap and its pointed tip, either blurring them (Gaussian-Flow and STG) or distorting their geometry (SplineGS). In contrast, ours is the only method that faithfully reconstructs both the cap's rounded geometry and sharp tip.

Finally, in \emph{flames}, the rapidly evolving flame and the moving hand holding the can present the greatest challenges. Gaussian-Flow, STG, and SplineGS collapse the flame into a saturated, blurry blob and smear the hand and can, whereas our method preserves more of the flame's tongue-like structure and reconstructs the hand and can with the highest sharpness (see zoom-ins).

\textbf{Quantitative Evaluation.}
Table~\ref{tab:all} summarizes the quantitative evaluation of our method in comparison with the other three SOTA methods: Gaussian-Flow, STG, and SplineGS on the five sequences, i.e., \emph{checker sphere}, \emph{soft contact}, \emph{cook spinach}, \emph{coffee martini}, and \emph{flames}.

Every method is trained with its own default iteration budget: Gaussian-Flow and STG use $30$k iterations on all five sequences; SplineGS uses $20$k on \emph{checker sphere}, \emph{cook spinach}, \emph{coffee martini} and \emph{flames}, and $40$k on \emph{soft contact}; our method uses $20$k on \emph{checker sphere}, $40$k on \emph{soft contact}, and $30$k on the three real captures.

Our method achieves the highest PSNR on all three real-world scenes, with a margin of $3.46$\,dB on \emph{flames}, as well as on the synthetic \emph{checker sphere} scene, and ranks second on \emph{soft contact}.

For SSIM, ours achieves the highest score on both synthetic scenes, i.e., \emph{checker sphere}, and \emph{soft contact}, as well as on~\emph{coffee martini} and \emph{flames}, and ranks second on \emph{cook spinach}. 

For LPIPS, it achieves the best score on both synthetic scenes (~\emph{checker sphere} and \emph{soft contact}) and on \emph{flames}, and ranks second on \emph{cook spinach} and \emph{coffee martini}, with only a small difference from the top-performing method in both cases.

Note that while our method ranks first or second across all tests and metrics, its performance is notably stronger on the synthetic scenes. This is expected, as our method is specifically designed to handle highly non-uniform dynamics. Our synthetic scenes were deliberately designed to exhibit such heterogeneous temporal dynamics, providing a controlled setting in which to evaluate the benefits of adaptive temporal allocation.

\textbf{Ablation Study.}
Table~\ref{tab:ablation_k} ablates the two design choices of our temporal parameterization: the spline order $K$ and the placement of the joints. For each order, we compare uniform spacing with our learnable non-uniform spacing.

Non-uniform spacing consistently outperforms (or matches) uniform spacing across all spline orders. Interestingly, increasing the spline order does not improve reconstruction quality. Instead, the lowest-order spline with non-uniform spacing achieves the best overall results, suggesting that adaptive joint placement is more important than spline order. Moreover, the performance gap between non-uniform and uniform spacing decreases as the spline order increases, as expected, since higher-order splines provide greater flexibility even with uniformly spaced joints.

\begin{table}[H]
\centering
\caption{Ablation on spline order $K$ with uniform and non-uniform joint spacing.
All metrics are averaged over five scenes (cut roasted beef, cook spinach, sear
steak, flame steak, coffee martini).}
\setlength{\tabcolsep}{11pt}
\begin{tabular}{llccc}
\hline
$K$ & Joints & PSNR$\uparrow$ & SSIM$\uparrow$ & LPIPS$\downarrow$ \\
\hline
\multirow{2}{*}{2} & Uniform     & 31.71 & 0.9459 & \textbf{0.0537} \\
                   & Non-uniform & \textbf{32.08} & \textbf{0.9481} & 0.0540 \\
\hline
\multirow{2}{*}{3} & Uniform     & 30.83 & 0.9441 & 0.0565 \\
                   & Non-uniform & \textbf{31.74} & \textbf{0.9469} & \textbf{0.0551} \\
\hline
\multirow{2}{*}{4} & Uniform     & \textbf{31.61} & \textbf{0.9454} & 0.0579 \\
                   & Non-uniform & \textbf{31.61} & 0.9446 & \textbf{0.0572} \\
\hline
\end{tabular}
\label{tab:ablation_k}
\end{table}

\section{Conclusions and limitations.}

We presented GaussianSpline, a dynamic 4D Gaussian Splatting method in which each Gaussian follows a piecewise cubic Bézier trajectory with a \emph{learnable, non-uniform} temporal partition. With a fixed segment budget, trajectories allocate temporal resolution where motion is fast and use larger intervals where it is slow. A differentiable reflection-based $G^1$ constraint ensures smooth trajectories without restricting local motion speed, and the method integrates directly into the standard 3DGS optimization and densification pipeline without auxiliary networks.

Our ablation study shows that non-uniform spacing consistently outperforms or matches uniform spacing across spline orders, while increasing the order provides little benefit (Table~\ref{tab:ablation_k}). GaussianSpline achieves the best overall results on \emph{checker sphere} and leads PSNR on real N3DV and Google Immersive scenes. 

Our approach has several limitations. First, while our spline representation is effective for a wide range of motion, it can struggle when the dynamics become highly transient. In \emph{Soft Contact}, for example, rapid collisions produce large accelerations over very short intervals that are difficult to capture accurately with a fixed, small number of spline segments. A promising direction is therefore adaptive segment \emph{insertion}, guided by local curvature or reconstruction residuals, to dynamically increase temporal resolution where needed.

Second, our trajectories are purely temporal and optimized independently for each Gaussian, without explicit spatial motion priors or coupling between neighboring primitives. As a result, rigid and articulated structures must be learned through individual Gaussian trajectories, while topological changes, such as the appearance and disappearance of geometry, are handled implicitly through opacity and densification.

Finally, our evaluation is limited to the publicly available N3DV scenes, one Google Immersive scene, and two synthetic sequences, with each baseline trained using its default optimization schedule. Broader evaluation is desirable, but diverse multi-view dynamic datasets with dense temporal sampling and accurate ground truth remain scarce. As more public datasets become available, evaluation across a wider range of motions and scene dynamics will provide a more comprehensive assessment.

\bibliographystyle{eg-alpha}
\bibliography{references.bib}

\end{document}